\documentstyle[12pt]{article}
\newcommand{\ba}[1]{\begin{eqnarray} \label{#1}}
\newcommand{\ea}{\end{eqnarray}}
\newcommand{\nn}{\nonumber}
\newcommand{\ban}{\begin{eqnarray*}}
\newcommand{\ean}{\end{eqnarray*}}
\def\half {\frac{1}{2}}
\def\ifmath#1{\relax\ifmmode #1\else $#1$\fi}
\def\ls#1{\ifmath{_{\lower1.5pt\hbox{$\scriptstyle #1$}}}}
\def\WT{\widetilde}

\def\half{{1\over 2}}
\def\sq2{\sqrt{2}}

\def\mchi{\mbox{$m_{\chi}$}}
\begin{document}
\begin{center}
{\Large \bf 
	On importance of dark matter for LHC physics%
\footnote{Talk given at the International symposium 
	``LHC physics and detectors'', Dubna, 28--30 June, 2000.}
\par }

\bigskip

{\large V.A.~Bednyakov \bigskip }

	Laboratory of Nuclear Problems,
	Joint Institute for Nuclear Research,
        Moscow region, 141980 Dubna, Russia;
	E-mail: bedny@nusun.jinr.ru
\end{center}

\begin{abstract} 
	I would like to attract attention of the LHC high-energy physics 
	community to non-accelerator, low-energy experiments, that are  
	also very sensitive to new physics. My example concerns search 
	for supersymmetric dark matter particles.
	It is shown that non-observation of the SUSY dark matter
	candidates with a high-accuracy detector can
	exclude large domains of the MSSM parameter space 
	and, in particular, can make especially desirable
	collider search for light SUSY charged Higgs boson.
\end{abstract} 

	A direct dark matter search for 
	neutralinos, lightest SUSY particles (LSP), is complementary
	to high energy searches for SUSY with colliders
\cite{DressNojiriRate}--\cite{9706509}.
	For example, colliders unable to prove that  
	the LSP is a stable particle.
	Such dark matter searches offer interesting 
	prospects for beating accelerators in discovery of 
	SUSY, particularly during	
	the coming years before the LHC enters into operation
\cite{h0001005}.

\begin{figure}[h!] 
\begin{minipage}[b]{0.50\textwidth}{
\begin{picture}(80,57)
\put(-6,-64){\includegraphics{Kepler.ps} }
\end{picture}
\caption{\small 
	Rotation curve of the solar system which falls off as
	$v = \sqrt{G^{}_{\rm N}M/r}$ 
	in accordance with Kepler's law. 
	AU is the Earth-Sun distance of $1.5\times 10^{13}$~cm}
\label{SolarRotationCurve}
}\end{minipage} \hfill
\begin{minipage}[b]{0.40\textwidth}{
\begin{picture}(80,57)
\put(-10,-16.5){\includegraphics{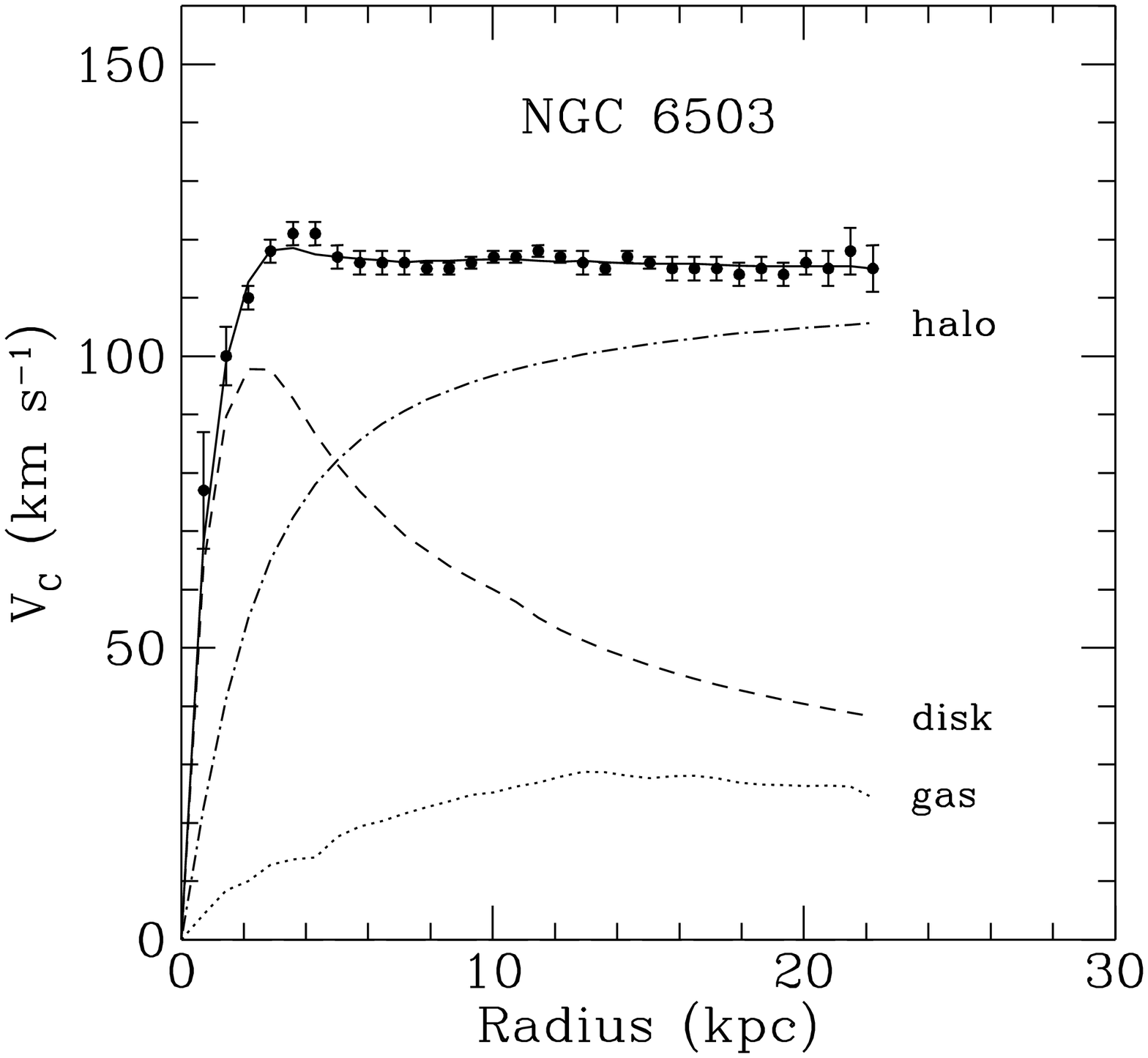} }
\end{picture}
\caption{\small Rotation curve of the spiral galaxy
	 NGC 6503 as established from radio 
	observations of hydrogen gas in the disk }
\label{SpiralRotationCurve}
}\end{minipage}
\end{figure}

	By definition, Galactic Dark Matter (DM) does not emit
	detectable amounts of electromagnetic radiation 
	and (only) gravitationally affects other, visible, celestial bodies. 
	The best evidence of this kind comes from the study 
	of galactic rotation curves,
	when one measures the velocity with which globular stellar
	clusters, gas clouds, or dwarf galaxies 
	orbit around their centers. 
	If the mass of these galaxies were
	concentrated in their visible parts, the orbital velocity at large
	radii $r$ should decrease as $1/\sqrt{r}$
(fig.~\ref{SolarRotationCurve}). 
	Instead, it remains approximately constant to the
	largest radius where it can be measured. 
	This implies that the total mass $M(r)$ felt by an object at a
	radius $r$ must increase linearly with $r$
(fig.~\ref{SpiralRotationCurve}). 
	Studies of this type imply that 90\% or more of the mass of
	large galaxies is dark
\cite{h9804231}.

\begin{figure}[h!]
\begin{minipage}[b]{0.60\textwidth}
{\begin{picture}(100,74)
\put(-7,-72){\includegraphics{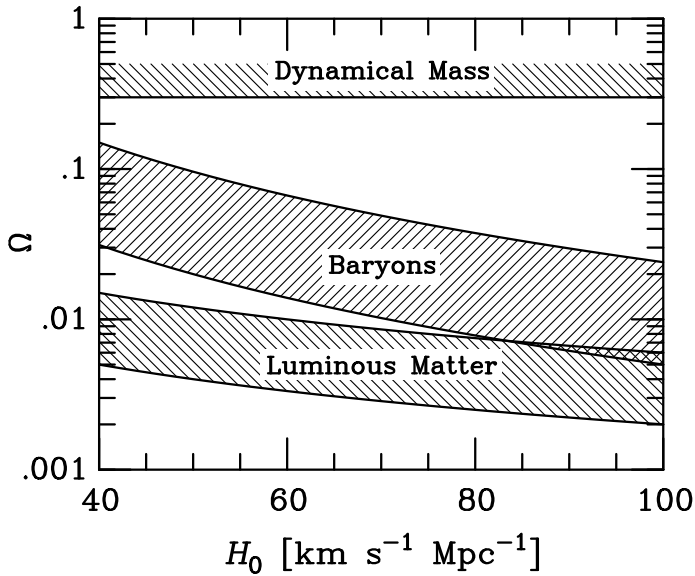} }
\end{picture}
}\end{minipage}
\hfill
\begin{minipage}[b]{0.37\textwidth}{
\caption{Most Dark Matter must be non--baryonic: \protect\\
	$\Omega=\Omega_{\mbox{\scriptsize baryon}}
	+\Omega_{\mbox{\scriptsize non-baryon}} \geq\! 0.3$, 
	with $\Omega_{\rm baryon} \leq 0.06$}
\label{Non-baryonic}
}\end{minipage}
\end{figure}

	The mass density averaged over the entire Universe is 
	usually expressed in units of critical density 
	$\rho_{\rm c} \approx 10^{-29}$g/cm$^3$, 
	the dimensionless ratio $\Omega \equiv \rho/\rho_{\rm{c}}= 1$ 
	corresponds to a flat Universe. 
	Analyses of galactic rotation curves imply $ \Omega \geq 0.1$.
	Studies of clusters and superclusters of galaxies through 
	gravitational lensing or through measurements of their X-ray 
	temperature, as well as studies of the large-scale streaming of 
	galaxies favor larger values of the total mass density of the 
	Universe $ \Omega \geq 0.3.$
	Finally, naturalness arguments and inflationary models prefer 
	$\Omega = 1.0$ to a high accuracy.
	The requirement that the Universe be at least 10 billion years old 
	implies $ \Omega h^2 \leq 1$, where $h = 0.65 \pm 0.15$ 
	is the present Hubble parameter in units of 100 km/(sec$\cdot$Mpc). 
	The total density of luminous matter only amounts 
	to less than 1\% of the critical density. 
	Analyses of Big Bang nucleosynthesis 
	determine the total baryonic density to lie in the range
	$0.01 \leq \Omega_{\rm baryon} h^2 \leq 0.015.$
	The upper bound implies $\Omega_{\rm baryon} \leq 0.06,$
	in obvious conflict with the lower bound $ \Omega \geq 0.3$. 
	Most Dark Matter must therefore be non--baryonic
(fig.~\ref{Non-baryonic}).
 
	Some sort of ``new physics'' seems to be required to describe 
	this exotic matter, beyond the particles described by the Standard
	Model of particle physics
\cite{h9804231}. 
	According to the best estimate, 
	the local density of this invisible matter amounts to about
$$
\rho_{\rm local}^{\rm DM} \simeq 0.3 \ {\rm GeV/cm}^3
\simeq 5 \cdot 10^{-25} {\rm g/cm}^3.
$$
	It is assumed to have a Maxwellian velocity distribution with mean
	$\bar v \simeq 300$~km/sec. 
	The local flux of DM particles $\chi$ is thus
$$ 
\Phi_{\rm local}^{\rm DM} \simeq \frac {100 \ {\rm GeV}} {\mchi}
\cdot 10^5 \ {\rm cm}^{-2} {\rm s}^{-1}.
$$
	This not-small-enough value is considered as a
	basis for direct search for dark matter particles.
	 
        A dark matter event is elastic scattering 
	of a relic DM neutralino from a target nucleus producing a nuclear 
	recoil which can be detected by a suitable detector
\cite{JuKaGr,h9908427,Superlight}.
 	The differential event rate 
	(per unit mass of the target nucleus) in respect to the recoil 
	energy is the subject of experimental measurements:
$$ 
\frac{dR}{dE_r} = {\Bigl[ N \frac{\rho_\chi}{m_\chi} \Bigr]}
	\displaystyle
   \int^{v_{\rm esc}}_{v_{\rm min}} dv f(v) v 
	{\frac{d\sigma}{dq^2}}(v,\! E_r),
$$
	where $q^2 = 2 M_A E_r$, $v_{\rm esc} \approx 600$~km/s,
	$\rho_{\chi} \approx 0.3~$GeV$/$cm$^{3}$, 
	$v_{\rm min}=\left(M_A E_r/2 M_{\rm red}^2\right)^{1/2}$,
 	$M_A$ is the mass of the target nucleus, and 
	$M_{\rm red}$ is the reduced mass.
        A typical nuclear recoil energy is $E_r\approx 10^{-6} m_{\chi}$. 
	The rate depends on the distribution of
        the DM neutralinos in the solar vicinity $f(v)$ and
        the cross section of neutralino-nucleus elastic scattering:
\ban
{v^2}{\frac{d\sigma}{dq^2}}(v,q^2) 
	&\propto& 
	\left[
	\underbrace{c_0^2({\cal C}_q, f_q)}_{\rm SUSY} 
	\underbrace{A^2\ {\cal F}_{\rm S}^2(q^2)}_{\rm scalar}
	\right.\\[0.1cm]  
	&+& \left.
    	\underbrace{a_0^2({\cal A}_{q}, \Delta q)}_{\rm SUSY}
	\underbrace{{\cal F}_{00}^2(q^2)}_{\rm spin}  + a_0 a_1 
	\underbrace{{\cal F}_{10}^2(q^2)}_{\rm spin}  +
	\underbrace{a_1^2({\cal A}_{q}, \Delta q)}_{\rm SUSY}
	\underbrace{{\cal F}_{11}^2(q^2)}_{\rm spin} 
    	\right].
\ean
	Here 
	$a^{}_{0,1}=\sum^p_q {\cal A}_{q} \Delta q 
	 \pm\sum^n_q {\cal A}_{q} \Delta q$, and 
	$c^{}_0 = \sum^{p,n}_q {\cal C}_{q} f_q $.
	The first term in brackets, which has $A^2$ enhancement,  
	corresponds to the so-called spin-independent or 
	scalar interaction, the other terms give 
	parametrization of the so-called spin-dependent interaction. 
	The nuclear structure
	presented by the scalar ${\cal F}_{\rm S}^2(q^2)$
	and spin ${\cal F}_{ij}^2(q^2)$ form factors 
	is factorized out of the nucleon structure
	(given via quark contributions 
	to the spin  $\Delta q$ and to the mass $f_q$ of the nucleon)
	and the SUSY contribution (${\cal C}_q$ and ${\cal A}_{q}$), 
	which enters into the calculations
	at the level of neutralino-quark effective low-energy 
	interaction via the Lagrangian:
$$
  L_{\rm eff} = 
	{\cal A}_{q}\cdot
      \bar\chi\gamma_\mu\gamma_5\chi\cdot
                \bar q\gamma^\mu\gamma_5 q +
    \frac{m_q}{M_{W}} \cdot{\cal C}_{q}\cdot\bar\chi\chi\cdot\bar q q
      \ + ... 
$$
        where the terms with vector and pseudoscalar quark currents are
        omitted being negligible in the case of non-relativistic
        DM neutralinos with typical velocities $v_\chi\approx 10^{-3} c$;
\ban
{\cal A}_{q} =
	&-&\frac{g^{2}}{4M_{W}^{2}}
	   \left[\frac{{\cal N}_{14}^2-{\cal N}_{13}^2}{2}T_3 
-\frac{M_{W}^2(\cos^{2}\theta_{q}\phi_{qL}^2+\sin^{2}\theta_{q}\phi_{qR}^2)}
      {m^{2}_{\tilde{q}1}-(m_\chi + m_q)^2}
	\right.\\ 
&-& 
\frac{M_{W}^2(\sin^{2}\theta_{q}\phi_{qL}^2+\cos^{2}\theta_{q}\phi_{qR}^2)}
{m^{2}_{\tilde{q}2} - (m_\chi + m_q)^2}\\
        &-& \frac{m_{q}^{2}}{4}P_{q}^{2}
	\left(\frac{1}{m^{2}_{\tilde{q}1}
		- (m_\chi + m_q)^2}
             + \frac{1}{m^{2}_{\tilde{q}2}
                - (m_\chi + m_q)^2}\right) \\
        &-& \frac{m_{q}}{2}\  M_{W}\  P_{q}\  \sin2\theta_{q}\
            T_3 ({\cal N}_{12} - \tan\theta_W {\cal N}_{11}) \nn \\
	&\times&
	\left.
\left( \frac{1}{m^{2}_{\tilde{q}1}- (m_\chi + m_q)^2}
    - \frac{1}{m^{2}_{\tilde{q}2} - (m_\chi + m_q)^2}\right)
	\right ]; 
\\[3pt]
 {\cal C}_{q} =
	&-&  \frac{g^2}{4} 
	\left[ 
	\frac{F_h}{m^2_{h}} h_q + \frac{F_H}{m^2_{H}} H_q 
	+ 
	\left(\frac{m_q}{4 M_W} P_{q}^{2} -
        \frac{M_W}{m_q} \phi_{qL}\ \phi_{qR}\right)
	\right.
\\&\times&
	\left(
	 \frac{\sin2\theta_{q}}{m^{2}_{\tilde{q}1} - (m_\chi + m_q)^2}
	-\frac{\sin2\theta_{q}}{m^{2}_{\tilde{q}2} - (m_\chi + m_q)^2}
	\right)
\\
	&+&  \left. 
	P_q \left(
	\frac{\cos^{2}\theta_{q}\ \phi_{qL} -
         \sin^{2}\theta_{q}\ \phi_{qR}}{m^{2}_{\tilde{q}1} - (m_\chi + m_q)^2}
         -\frac{\cos^{2}\theta_{q}\ \phi_{qR} -
     \sin^{2}\theta_{q}\ \phi_{qL}}{m^{2}_{\tilde{q}2} - (m_\chi +
	m_q)^2}\right)
\right].
\ean
\ba{Fq1}
        F_{h} &=& ({\cal N}_{12} - {\cal N}_{11}\tan\theta_W)
        ({\cal N}_{14}\cos\alpha_H + {\cal N}_{13}\sin\alpha_H), \nn \\
        F_{H} &=& ({\cal N}_{12} - {\cal N}_{11}\tan\theta_W)
        ({\cal N}_{14}\sin\alpha_H - {\cal N}_{13}\cos\alpha_H),\nn \\
h_q &=&\bigl(\frac{1}{2}+T_3\bigr)\frac{\cos\alpha_H}{\sin\beta}
          - \bigl(\frac{1}{2}-T_3\bigr)\frac{\sin\alpha_H}{\cos\beta},\nn \\
H_q &=& \bigl(\frac{1}{2}+T_3\bigr)\frac{\sin\alpha_H}{\sin\beta}
      + \bigl(\frac{1}{2}-T_3\bigr)\frac{\cos\alpha_H}{\cos\beta}, \nn \\
\phi_{qL} &=& {\cal N}_{12} T_3 + {\cal N}_{11}(Q -T_3)\tan\theta_{W},\nn
\qquad
\phi_{qR} = \tan\theta_{W}\  Q\  {\cal N}_{11}, \nn \\
P_{q} &=&  \bigl(\frac{1}{2}+T_3\bigr) \frac{{\cal N}_{14}}{\sin\beta}
          + \bigl(\frac{1}{2}-T_3\bigr) \frac{{\cal N}_{13}}{\cos\beta}. \nn
\ea

	In this paper the MSSM parameter space is explored at the weak
	scale, when any constraints following from the unification assumptions 
	are completely relaxed.
	On the other side, restrictions from the age of the Universe, 
	accelerator SUSY searches, rare FCNC $b\to s\gamma$ decay, etc
	are respected
\cite{kolb}--\cite{roskane}.
	Therefore, the MSSM parameter space is determined 
	by entries of the mass matrices of neutralinos, charginos, 
	Higgs bosons, sleptons and squarks.
	All relevant mass matrices are given below.
	The one-generation squark and slepton mass matrices  
	have the form
\cite{haka}:
\ba{Squarks} \nn
        M^2_{\tilde t} &=& \left[ \matrix{%
        M^2_{\widetilde Q}+ m^2_t+ m^2_Z(\half-\frac23 s^2_W)\cos2\beta
        & m_t(A_t-\mu\cot\beta) \cr
        m_t(A_t-\mu\cot\beta)
        &M^2_{\widetilde U}+ m^2_t+ m^2_Z \frac23 s^2_W\cos2\beta \cr}
      \right], \\
\nn
        M^2_{\tilde b} &=& \left[ \matrix{%
    M^2_{\widetilde Q}+ m^2_b- m^2_Z(\half-\frac13 s^2_W)\cos2\beta
      & m_b(A_b-\mu\tan\beta) \cr
    m_b(A_b-\mu\tan\beta)
      &M^2_{\widetilde D}+ m^2_b - m^2_Z \frac13 s^2_W\cos2\beta \cr }
      \right],
\\ \nn
  M^2_{\tilde\nu} &=& M^2_{\widetilde L} + \half m^2_Z\cos2\beta, \\
\nn 
  M^2_{\tilde \tau}  &=& \left[ 
                        \begin{array}{cc}
    M^2_{\widetilde L}+ m^2_\tau- m^2_Z(\half- s^2_W)\cos2\beta
      & m_\tau(A_\tau-\mu\tan\beta) \\
    m_\tau(A_\tau-\mu\tan\beta)
      &M^2_{\widetilde E}+ m^2_\tau- m^2_Z s^2_W\cos2\beta 
                        \end{array}
      \right]
\ea
	where $s^2_W\equiv\sin^2\theta_W$ and
	$\tan\beta \equiv \langle {H^0_2} \rangle / \langle {H^0_1}\rangle$.
	In the $\WT W^+$--$\WT H^+$ basis, the chargino mass matrix is 
$$
  X = \pmatrix{ M_2 &\sqrt 2 m\ls W \sin\beta \cr
       \sqrt 2 m\ls W \cos\beta &\mu \cr}.  
$$
	Two unitary $2\times 2$ matrices $U$ and $V$ are required
	to diagonalize the chargino mass-squared matrix
$
{\cal M}_{\WT\chi^+}^2=VX^\dagger XV^{-1}=U^\ast XX^\dagger
(U^\ast)^{-1}.
$
	The two mass eigenstates are denoted by $\widetilde\chi^+_1$
	and $\widetilde\chi^+_2$.
	In the $\WT B$--$\WT W^3$--$\WT H^0_1$--$\WT H^0_2$ basis, the
	neutralino Majorana mass matrix is
$$
  Y = \pmatrix{%
    M_1 &0  &-m_Zc_\beta s_W &m_Zs_\beta s_W \cr
    0   &M_2 &m_Z c_\beta c\ls W &-m_Z s_\beta c\ls W \cr
    -m_Z c_\beta s\ls W &m_Z c_\beta c\ls W &0 &-\mu \cr
    m_Z s_\beta s\ls W &-m_Z s_\beta c\ls W &-\mu &0 \cr }\,,
$$
	where $s_\beta = \sin\beta$, $c_\beta=\cos\beta$, etc.  
	A $4\times 4$ unitary matrix ${\cal N}$ is required to diagonalize 
	the neutralino mass matrix
	${\cal M}_{\WT\chi^0}={\cal N}^\ast Y{\cal N}^{-1} $
	where the diagonal elements of ${\cal M}_{\WT\chi^0}$ can 
	be either positive or negative.
	The CP-even Higgs mass matrix has the form
\cite{Higgses}
\begin{eqnarray*}
       \left(\begin{array}{cc}
               H_{11} & H_{12}\\  
               H_{12} & H_{22} 
          \end{array}\right)
	&=&
	\half \left(\begin{array}{cc}
                 \tan\beta & -1 \\   -1 & \cot\beta 
         \end{array}\right) 
                  M_A^2\sin 2\beta 
\\ \nn
      &+&\half \left(\begin{array}{cc}
                  \cot\beta & -1 \\   -1 & \tan\beta 
           \end{array}\right) m_Z^2\sin 2\beta 
       +\frac{3g_2^2}{16\pi^2m_W^2}
         \left(\begin{array}{cc}
               \Delta_{11} & \Delta_{12}\\  
               \Delta_{12} & \Delta_{22} 
          \end{array}\right);
\end{eqnarray*}
\ba{d11}
\nn
 \Delta_{11} &=& \frac{m^4_b}{c^2_\beta}
 (  \ln{\frac{m^2_{\tilde{b}_1} m^2_{\tilde{b}_2}} {m^4_b}}
  + \frac{2 {A}_b( {A}_b-\mu \tan\beta)}
         {m^2_{\tilde{b}_1}-m^2_{\tilde{b}_2}}
   \ln{ \frac{m^2_{\tilde{b}_1}} {m^2_{\tilde{b}_2}} }  ) \\
\nn   &+& \frac{m^4_b}{c^2_\beta}
( \frac{ {A}_b( {A}_b-\mu \tan\beta)}
      {m^2_{\tilde{b}_1}-m^2_{\tilde{b}_2}})^2
 g(m^2_{\tilde{b}_1}, m^2_{\tilde{b}_2} ) 
	+ \frac{m^4_t}{s^2_\beta}
 ( \frac{ \mu( {A}_t-\frac{\mu}{\tan\beta})} 
        {m^2_{\tilde{t}_1}-m^2_{\tilde{t}_2}})^2
 g(m^2_{\tilde{t}_1}, m^2_{\tilde{t}_2} ).
\\ \nn
 \Delta_{22} &=& \frac{m^4_t}{s^2_\beta}
 ( \ln{ \frac{ m^2_{\tilde{t}_1} m^2_{\tilde{t}_2} }
             { m^4_t                               }      }
     + \frac{ 2 {A}_t( {A}_t-\frac{\mu}{\tan\beta}) }
             { m^2_{\tilde{t}_1}-m^2_{\tilde{t}_2}        }
             \ln{ \frac{ m^2_{\tilde{t}_1} }
                       { m^2_{\tilde{t}_2} }              }
 )      \\
\nn      &+&  \frac{m^4_t}{s^2_\beta}
    ( \frac{ {A}_t ( {A}_t-\frac{\mu}{\tan\beta} ) }
          { m^2_{\tilde{t}_1}-m^2_{\tilde{t}_2}          }
   )^2  g( m^2_{\tilde{t}_1},m^2_{\tilde{t}_2} ) 
+\frac{m^4_b}{c^2_\beta}
   ( \frac{ \mu ( {A}_b-\mu\tan\beta )       }
          {m^2_{\tilde{b}_1} - m^2_{\tilde{b}_2} }
   )^2 g( m^2_{\tilde{b}_1},m^2_{\tilde{b}_2} ).
\\ \nn
  \Delta_{12} &=& \Delta_{21} = \frac{m^4_t}{s^2_\beta}
     \frac{\mu ( {A}_t-\frac{\mu}{\tan\beta}) }
          { m^2_{\tilde{t}_1}-m^2_{\tilde{t}_2} }
  (  \ln{ \frac{ m^2_{\tilde{t}_1} }
               { m^2_{\tilde{t}_2} }
        }
      + \frac{  {A}_t( {A}_t-\frac{\mu}{\tan\beta})}
             {  m^2_{\tilde{t}_1} - m^2_{\tilde{t}_2}    }
      g(  m^2_{\tilde{t}_1}, m^2_{\tilde{t}_2}   )
  )\\
\nn    &+& \frac{m^4_b}{c^2_\beta}
     \frac{\mu ( {A}_b- \mu\tan\beta) }
          { m^2_{\tilde{b}_1}-m^2_{\tilde{b}_2} }
  (  \ln{ \frac{ m^2_{\tilde{b}_1} }
               { m^2_{\tilde{b}_2} }
        }
      + \frac{ {A}_b( {A}_b-\mu\tan\beta) }
             {  m^2_{\tilde{b}_1} - m^2_{\tilde{b}_2}    }
      g(  m^2_{\tilde{b}_1}, m^2_{\tilde{b}_2}   )
  ).
\ea
	Here $c^2_\beta = \cos^2\!\beta$, $s^2_\beta = \sin^2\!\beta$ and
	$\displaystyle g(m_1^2,m_2^2) =  2 - \frac{m_1^2+m_2^2}{m_1^2-m_2^2}
        \ln{ \frac{m_1^2}{m_2^2} }.$
	Neutral CP-even Higgs eigenvalues are	
$$
m_{H,h}^2 = \half \Bigl\{ H_{11}+H_{22}
             \pm \sqrt{(H_{11}+H_{22})^2 - 4(H_{11}H_{22}-H_{12}^2)} \Bigr\},
$$
	The mixing angle $\alpha_H^{}$ is obtained from
$$
   \sin 2\alpha_H^{}= \frac{2H^2_{12}}{m^2_{H^0_1} - m^2_{H^0_2}},
\qquad
   \cos 2\alpha_H^{}= \frac{H^2_{11}-H^2_{22}}{m^2_{H^0_1} - m^2_{H^0_2}}.
$$ 
	The mass of the charged Higgs boson is given by
  	$m^2_{\rm CH} = m^2_W + M^2_A  + 
\frac{3g_2^2}{16\pi^2m_W^2}
\Delta_{\rm ch}$
\cite{Higgses}.%
\enlargethispage{0.5\baselineskip}

	Therefore free parameters are 
	$\tan\beta$, the ratio of neutral Higgs boson vacuum expectation 
	values; 
	$\mu$, the bilinear Higgs parameter of the superpotential;
	$M_1$ and $M_2$, soft gaugino masses; 
	$M_A$, the CP-odd Higgs mass; 
	$m^2_{\WT Q}$, $m^2_{\WT U}$ and $m^2_{\WT D}$, 
	squark mass parameters squared for the 1st and 2nd generation;        
	$m^2_{\WT L}$ and $m^2_{\WT E}$,  
	slepton mass parameters squared for the 1st and 2nd generation;
	$m^2_{\WT Q_3}$, $m^2_{\WT T}$ and $m^2_{\WT B}$, 
	3rd generation squark mass parameters squared; 
	$m^2_{\WT L_3}$ and $m^2_{\WT \tau}$, 
	3rd generation slepton mass parameters squared; 
	$A_t$, $A_b$ and $A_\tau$, soft trilinear 
	couplings for the 3rd generation.
	With these parameters one completely determines the MSSM spectrum
	and the coupling constants.
	Following 
\cite{h0007202}--\cite{h0005234} we assume that squarks are degenerate. 
	Bounds on flavor-changing neutral currents 
	imply that squarks with equal gauge
	quantum numbers must be close in mass. 
	Therefore, for the sfermion mass parameters we used the relations 
	$m^2_{\widetilde U_{}} = m^2_{\widetilde D_{}} 
	= m^2_{\widetilde Q_{}}$, 
	$m^2_{\widetilde E_{}} = m^2_{L}$, 
	$m^2_{\widetilde T} = m^2_{\widetilde B} = m^2_{Q_3}$,  
	$m^2_{\widetilde E_{3}} = m^2_{L_3}$.
	The parameters $A_b$ and $A_\tau$ are fixed to be zero.

   	The present lifetime of the Universe implies an upper limit 
	on the expansion rate and correspondingly on the total 
	relic abundance.
 	Assuming that the neutralinos form a dominant part of
    	the dark matter in the Universe
\cite{DressNojiriRate,Superlight,9801246} 
	one obtains a lower limit on the neutralino relic density.
	In this analysis the cosmological constraint 
	$0.025 < \Omega_\chi h^2_0<1$ was implemented%
\footnote{Recently exciting evidence for a flat and 
	accelerating universe has been obtained 
\cite{a0004404,a0005124}, which results in a more stringent cosmological
	constraint $0.1 < \Omega_\chi h^2_0< 0.3$.
	This new constraint does not affect the main result of the paper.}.
 	The neutralino mass density parameter 
	$\Omega_{\chi} h^2_0$  was calculated by the standard
   	procedure on the basis of the approximate formula
\cite{drno}. 
	All channels of the $\chi-\chi$ annihilation are included.
   	Since neutralinos are mixtures of gauginos and
   	higgsinos, annihilation can occur both via
   	s-channel exchange of the $Z^0$ and Higgs bosons and
   	t-channel exchange of a scalar particle
\cite{roskane,drno,relic}.
        Another stringent constraint is imposed by the 
	branching ratio of the $b\to s\gamma$ decay, 
	measured by the CLEO collaboration to be 
	$1.0 \times 10^{-4} < {\rm B}(b\to s \gamma) < 4.2 \times 10^{-4}$.
        In the MSSM this flavor-changing neutral-current 
        process receives contributions from $H^\pm$--$t$,
        $\tilde{\chi}^\pm$--$\tilde{t}$ and $\tilde{g}$--$\tilde{q}$ loops
        in addition to the standard model $W$--$t$ loop.
	These also strongly restrict the parameter space
\cite{BerBorMasRi}.

	The masses of the supersymmetric particles are constrained 
	by the results from the high energy colliders.
	This  imposes constraints on the parameter space of the MSSM.
	The following experimental restrictions are used
\cite{PDG}:
$ M_{\tilde{\chi}^+_{2}} \geq 65$~GeV for the light chargino,
$ M_{\tilde{\chi}^+_{1}} \geq 99$~GeV for the heavy chargino,
$ M_{\tilde{\chi}^0_{1,2,3}} \geq 45, 76, 127$~GeV for non-LSP
					neutralinos,
$ M_{\tilde{\nu}}      \geq 43$~GeV for sneutrinos,
$ M_{\tilde{e}_R}      \geq 70$~GeV for selectrons,
$ M_{\tilde{q}}       \geq 210$~GeV  for squarks,
$ M_{\tilde{t}_1}      \geq 85$~GeV  for light top-squark,   
$ M_{H^0}              \geq 79$~GeV  for neutral Higgs bosons,
$ M_{\rm CH}           \geq 70$~GeV  for charged Higgs boson.

	In the numerical analysis a trial set of MSSM
	parameters is picked up randomly from the following intervals: 
\begin{eqnarray*}
&-1{\rm ~TeV} < M_1 < 1{\rm ~TeV}, \quad
-2{\rm ~TeV} < M_2, \mu, A_t < 2{\rm ~TeV},&  \\ \label{Scan}
&1 < \tan\beta < 50, \quad
60{\rm ~GeV} < M_A < 1000{\rm ~GeV},&\\
&10{\rm ~GeV}^2 < m^2_{Q_{}}, 
m^2_{L}, m^2_{Q_3}, m^2_{L_3}<10^6{\rm ~GeV}^2.&
\end{eqnarray*}
	For each trial set the MSSM particle masses and other
	observables are evaluated and compared with the 
	restrictions and constraints discussed above.
	If all constraints are successfully passed,
	the so-called total event rate $R$ 
	integrated over recoil energies is calculated
\cite{h9908427}.

\begin{figure}[h!] 
\begin{picture}(100,80)
\put(2,-110){\includegraphics{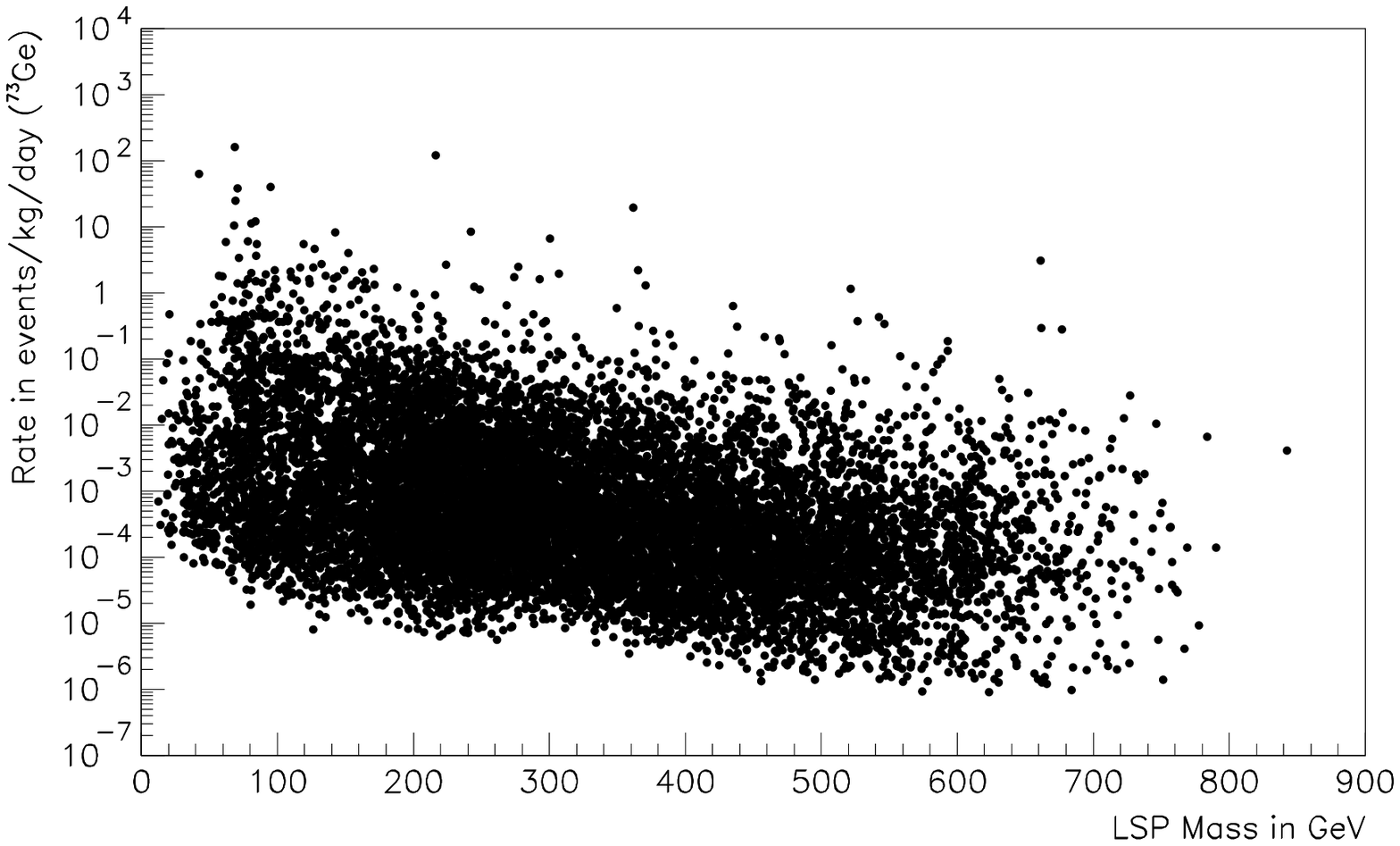}}
\end{picture} 
\caption{\small
	Total event rate in $^{73}$Ge versus the mass of LSP.
 	The lower bound decreases with increasing the mass of the 
	LSP and reaches the absolute minimum of about $10^{-6}$ 
	events/day/kg} 
\label{Rate-scan}
\begin{picture}(100,103)
\put(10,-32){\includegraphics{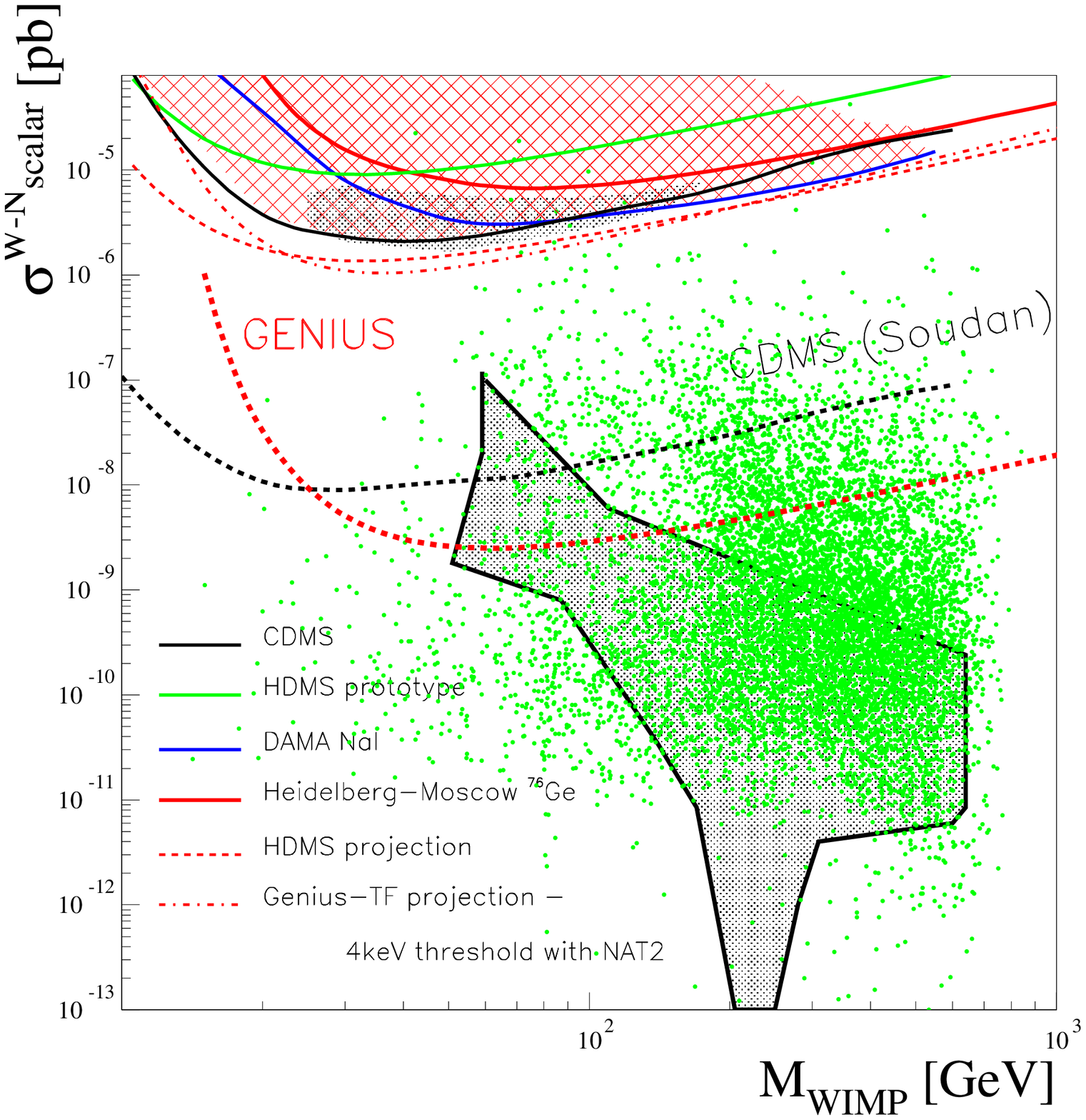}}
\end{picture}
\caption{\small 
	WIMP-nucleon cross section limits 
	for scalar interactions as a function of the WIMP mass. 
	Points are calculated without extra constraints.
	The filled area is obtained in 
\protect\cite{h0001005}.
	Some experimental results and expectations are also given
\protect\cite{GENIUS,NECRESST,HDMS}}
\label{WIMP-scan}
\vspace*{-0.5\baselineskip}
\end{figure}

    	The results of the scanning procedure are presented in
fig.~\ref{Rate-scan} as scatter plots.
	The main feature is a lower bound for the total event rate $R$. 
	An absolute minimum value of about $10^{-6}$ events/day/kg 
	in a $^{73}$Ge detector is obtained in the above-mentioned
	domain of the MSSM parameter space.
	There is a clear increase (up to one order of magnitude) 
	in the lower bound only with $\tan\beta$.
	In all other cases there is a decrease in the lower bound; 
	the decrease is the sharpest with $|\mu|$, $M^{}_A$ 
	(about 5 orders of magnitude) and $M^2_{Q_{}}$
	and with the squark mass $M^{}_{\WT q}$,  
	heavy chargino mass $M_{\tilde{\chi}^+_{1}}$
	and charged Higgs boson mass $M_{\rm CH}$.

	For comparison of the results obtained with sensitivities of different
	dark matter experiments the total cross section for relic 
	neutralino scalar elastic scattering 
	on the nucleon was also calculated and presented in 
fig.~\ref{WIMP-scan}.

\begin{figure}[th!] 
\begin{picture}(100,106)
\put(20,-2){\includegraphics{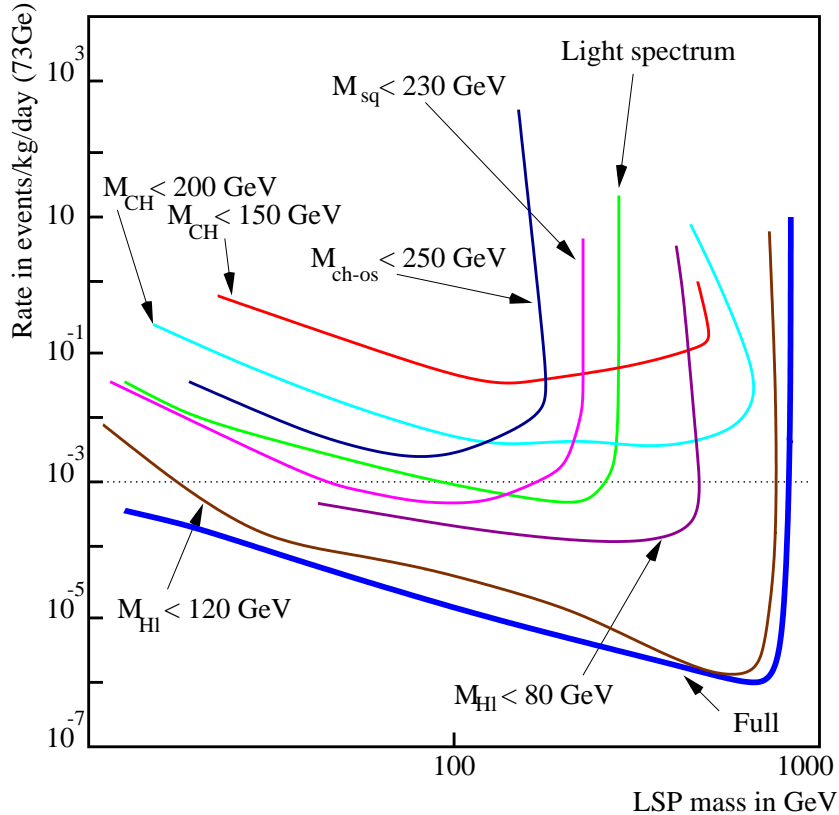}}
\end{picture}
\caption{\small Different lower bounds for the total  event rate in 
	$^{73}$Ge (events/day/kg)  versus the LSP mass (GeV).
	Here M$^{}_{\rm sq,\ CH,\ Hl}$ denote masses of the squark,
	the charged Higgs boson and the light neutral CP-even Higgs boson 
	respectively.
	The heavy chargino mass is denoted as M$^{}_{\rm ch-os}$.
	"Full" corresponds to the lower bound obtained 
	from the main (unconstrained) scan, and
	"Light spectrum" denotes the lower bound for $R$
	obtained with all sfermion masses lighter than about 300 GeV.
	The horizontal dotted line represents the expected sensitivity
	for direct dark matter detection with GENIUS
\protect\cite{GENIUS,NECRESST}}
\label{LowBounds}
\end{figure}

\begin{figure}[th] 
\begin{minipage}[b]{0.50\textwidth}
\begin{picture}(50,90)
\put(-43,-175){\includegraphics{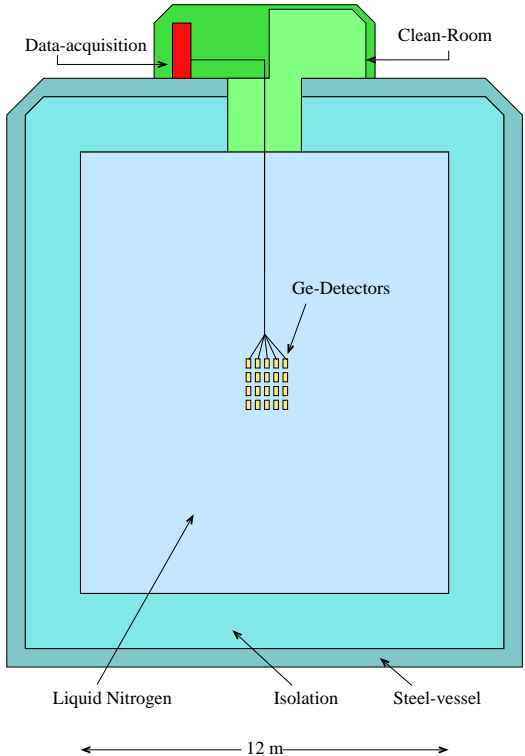}}
\end{picture}
\end{minipage}	\hfill
\begin{minipage}[b]{0.47\textwidth}{
\caption{\small 
	Schematic view of the GENIUS experiment:
	there are 300 enriched $^{76}$Ge detectors (1 ton) in 
	a liquid nitrogen shielding 
\protect\cite{GENIUS} 	\protect\\
}\label{GENIUS-setup}
}\end{minipage}
\end{figure}

	The variation of the lower bound for the event rate
	with the MSSM parameters and SUSY particle masses 
	allows one to consider prospects for dark matter search 
	under specific assumptions concerning these parameters and masses.
	To this end a number of extra scans with extra limitations
	on the single squark mass (M$^{}_{\rm sq}< 250,\ 230$~GeV), 
	light neutral CP-even Higgs boson mass 
	(M$^{}_{\rm Hl}< 80, \ 100, \ 120$~GeV), 
	charged Higgs boson mass (M$^{}_{\rm CH}<150, \ 200$~GeV) 
	and heavy chargino mass (M$^{}_{\rm ch-os}< 250$~GeV) were performed. 
	The case with light masses of all superpartners 
	(less than 300--400 GeV) was also considered.
	All corresponding curves together with the absolute
	lower bound from the unconstrained scan are depicted in 
Fig.~\ref{LowBounds}.
\enlargethispage{0.5\baselineskip}
	A restriction that the single (light) squark mass is small
	(M$^{}_{\rm sq}< 230$~GeV) as well as another assumption that all 
	sferminos masses do not exceed 300--400 GeV put upper limits on the 
	mass of the LSP and do not permit $R$ to drop very deeply as 
	the mass of the LSP increases.
	In both cases the lower bound for the
	rate is established for all allowed masses of the LSP
	at a level of $10^{-3}$ events/kg/day.     
	This value for the event rate is considered as an optimistic 
	sensitivity expectation for future high-accuracy 
	detectors of dark matter like GENIUS  
(Fig.~\ref{GENIUS-setup}).
	The same lower bound is obtained under the assumption that both 
	charginos are light (M$^{}_{\rm ch-os}< 250$~GeV).

\begin{figure}[th!] 
\begin{picture}(100,83)
\put(12,0){\includegraphics{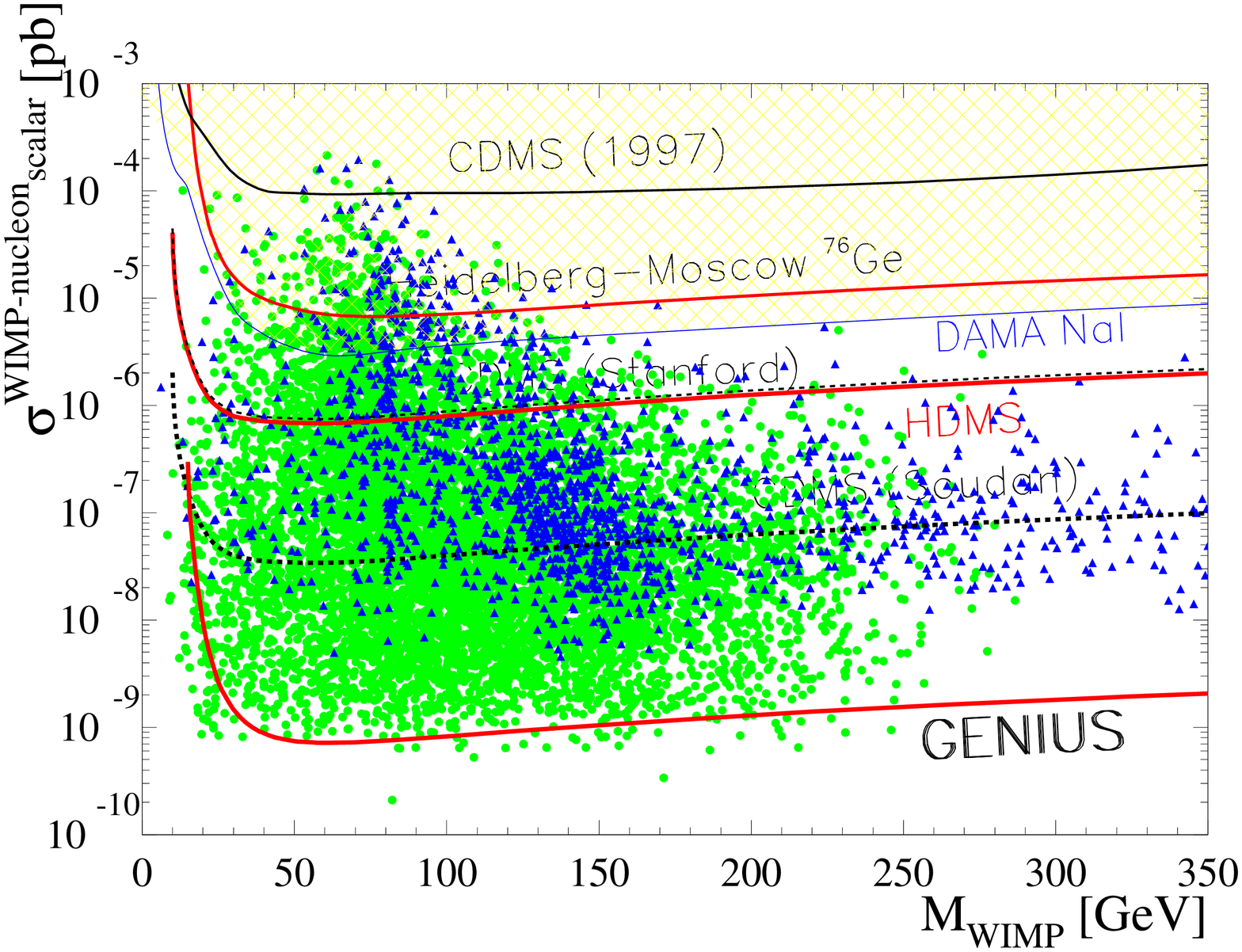}}
\end{picture}
\caption{\small 
	WIMP-nucleon cross section limits
	for scalar interactions as a function of the WIMP mass.
	Filled circles present calculations with the light SUSY spectrum. 
	Filled triangles give the cross section
	on the assumption that M$^{}_{\rm CH}<200$~GeV}
\label{WIMP-nucleon1}
\begin{picture}(100,87)
\put(12,-3){\includegraphics{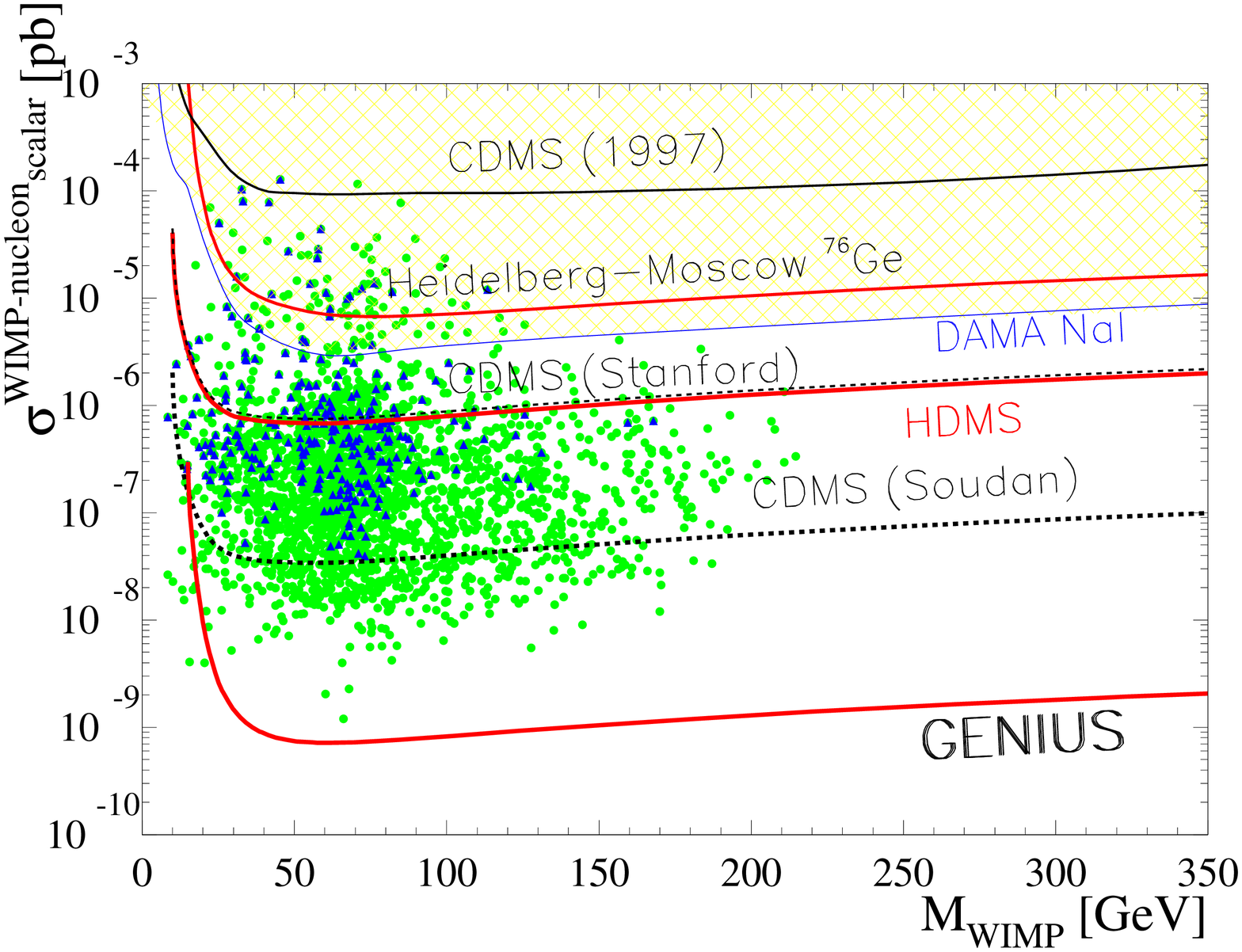}}
\end{picture}
\caption{\small 
	WIMP-nucleon cross section limits
	for scalar interactions as a function of the WIMP mass.
	Filled circles present our calculations
	with the light SUSY spectrum and $\tan\beta>20$. 
	Filled triangles give the same, but for $\tan\beta>40$}
\label{WIMP-nucleon2} 
\end{figure}

	It is seen that the mass of the light neutral CP-even 
	Higgs boson M$^{}_{\rm Hl}$
	has unfortunately a very poor restrictive potential
	(see, for example, the curve with M$^{}_{\rm Hl}< 120$~GeV,
	the curve with M$^{}_{\rm Hl}< 80$~GeV is already excluded). 
	The situation looks most promising with
	the mass of the charged Higgs boson. 
	When the charged Higgs boson mass is (relatively) small 
	the other masses of 
	CP-even and CP-odd Higgs bosons are also restricted from above.
	Therefore, couplings of the scalar neutralino-quark 
	interaction, which contain 
	$m^{-2}_{H,h}$-terms, are not suppressed
	enough and the rate cannot decrease significantly.
	The lower bound of the rate increases when the 
	mass M$^{}_{\rm CH}$ decreases 
	and for M$^{}_{\rm CH}<200\,$GeV (150~GeV) reaches
	the values of about $10^{-2}$ ($10^{-1}$) events/kg/day 
	practically for all allowed masses of the LSP.
	As can be seen from 
Fig.~\ref{WIMP-nucleon1}, these values can be reached not only with  
	GENIUS, but also 
	with some other near-future direct dark matter detectors
\cite{HDMS}.
	Filled circles in 
Fig.~\ref{WIMP-nucleon1} give the scalar cross sections calculated 
	under the assumption that the SUSY spectrum is light.
	Filled triangles give the cross section obtained
	with the charged Higgs boson mass restriction 
	M$^{}_{\rm CH}<200$~GeV. 
	If it happens, for instance,  that 
	either the SUSY spectrum is light indeed or 
	the charged Higgs boson mass really does not exceed 200~GeV,
	in both cases at least 
	the GENIUS experiment should detect a dark matter signal.
	If one considers a more complicated condition and 
	assumes that the SUSY spectrum is 
	light and simultaneously 
	that $\tan\beta$ is quite large, 
	then not only GENIUS, but also CDMS and HDMS
\cite{NECRESST, HDMS} will have very good
	prospects for detecting a dark matter signal.
	This situation is illustrated in
Fig.~\ref{WIMP-nucleon2}, where, besides
	cross section limits for the WIMP-nucleon scalar
	interactions for different experiments, 
	calculations for the case of a light SUSY spectrum
	with extra assumptions of $\tan\beta>20$ (filled circles)
	and $\tan\beta>40$ (filled triangles) are given.

\enlargethispage{0.5\baselineskip}
	Therefore, the correlations between the lower limit
	for the event rate $R$\ and some masses of SUSY particles
	allow good prospects for direct dark matter detection
	with next-generation detectors.
	The prospects could be brighter
	if collider searches would be able to restrict 
	the mass of the charged Higgs boson at a level of about 200 GeV 
	(light Higgs sector). 
	The observation, due to its importance for 
	dark matter detection, could serve as a stimulus for extra
	efforts to search for charged Higgs boson with colliders.
	Considered together, both these experiments, 
	collider search for charged Higgs boson and 
	dark matter search for SUSY LSP,   
	become very decisive for a verification of SUSY models. 

	On the contrary, 
	non-observation of any dark matter signal
	with very sensitive dark matter detectors, in accordance with 
Fig.~\ref{LowBounds}--\ref{WIMP-nucleon2}, would exclude, for example, 
	a SUSY spectrum with masses lighter than 300--400 GeV as well as 
	a light SUSY spectrum with large $\tan\beta$
(Fig.~\ref{WIMP-nucleon2}), charginos with masses smaller than 250 GeV 
(Fig.~\ref{LowBounds}), charged Higgs boson with M$^{}_{\rm CH}<200$~GeV,
	and therefore the entire light Higgs sector
	in the MSSM
(Fig.~\ref{WIMP-nucleon1}). 	
	
	The latter case is in particular interesting, because
	if the light charged Higgs boson is excluded by GENIUS, 
	then either it will be rather unpromising
	to search for it with colliders, or any positive result 
	of a collider search
	brings strong contradictions in the MSSM approach 
	to dark matter detections and/or collider SUSY searches.

\smallskip
 	These results may be considered as 
	a good example of the complementarity 
	of modern accelerator and non-accelerator 
	experiments looking for SUSY and other new physical phenomena.
\smallskip

	I would like to thank the organizers of the
	International symposium ``LHC physics and detectors''
	for the invitation to give this talk.
	I am grateful to Prof. H.V. Klapdor-Kleingrothaus
	for fruitful collaboration on the subject.	
	The investigation was supported in part 
	by RFBR Grant 00--02--17587.

\small 

\end{document}